\documentclass{aa}  
\usepackage{keyval}
\usepackage{amsmath}
\usepackage{multirow}
\usepackage{booktabs}
\usepackage{graphicx}
\usepackage{txfonts}
\usepackage{array}
\usepackage{siunitx}
\usepackage{lipsum}
\usepackage[mathscr]{euscript}
\usepackage{subcaption}         
\usepackage{lscape}             
\usepackage{placeins}           
                                
\usepackage[colorlinks=true,
            linkcolor=blue,
            citecolor=blue,
            urlcolor=blue]{hyperref}
\usepackage{xcolor}

\begin{document}

   \title{A self-supervised machine learning approach to identify instrumental systematics in 21cm Cosmology}

%
%
%

   \author{Ajith Sampath\inst{1}\corrauth{ajith.sampath@unige.ch}}        
   \institute{ Département de Physique Théorique and Center for Astroparticle Physics, Université de Genève, 24 quai Ernest-Ansermet, CH1211 Genève 4, Switzerland}

   \date{}

  \abstract
  {The primordial sound waves in the baryon-photon plasma are imprinted on the large-scale galaxy distribution as Baryon Acoustic Oscillations (BAO), serving as a cosmological standard ruler to constrain dark energy and the Hubble parameter. The upcoming 21cm intensity-mapping arrays are designed to map the 21cm emission across large volumes and measure the BAO scale. However, radio interferometers suffer from instrumental systematics that can hinder detection of the cosmological 21cm signal by leaking bright foregrounds into the signal-dominated modes.}
  {This work aims to demonstrate that a self-supervised model can predict the instrument systematic state (here, the amplitude of a spectral ripple in the beam width) in interferometric visibilities and potentially disentangle the instrument from the sky without a forward model.}
  {Based on the fact that an instrument stays the same as the sky rotates in a drift-scan observation, and that the same sky is observed under different instrument states, we train a convolutional autoencoder to compress the information in the \textit{simulated} visibilities and separate this information into instrument-only and sky-only latent vectors.}
  {The result shows excellent prediction of the unknown instrument systematic for 15 held-out instrument states with $R^2 = 0.9975$ with a linear probe, against $R^2 = -3.55$ for a random-encoder baseline. The instrument-latent carries no detectable sky signal, as measured with a total-power sky proxy, for any held-out test data. The sky-latent vector and instrument-latent vector show partial separation but not complete disentanglement, identifying sky-latent purity as the primary open problem.}
  {The results show the ability of the model to recognize the systematic and show a promising path towards an ML-based foreground cleaning pipeline. Full instrument-sky disentanglement constitutes the natural next step.}

   \maketitle

  \nolinenumbers

\section{Introduction}
Current and next generation cosmological surveys like DES \citep{des_collaboration_dark_2026}, DESI \citep{abdul_karim_desi_2025}, Rubin LSST \citep{ivezic_lsst_2019}, Euclid \citep{mellier_euclid_2025} and Roman \citep{spergel_wide-field_2015}, provide or promise tighter constraints on cosmological parameters including the dark energy equation of state parameters and Hubble parameter $H(z)$ by measuring the Baryon Acoustic Oscillations (BAO) - the primordial sound waves - imprinted in the large scale galaxy distribution. Another independent approach to solving the mystery of dark energy and resolving the existing tensions in the $\Lambda$CDM model is a technique known as intensity mapping (IM). IM uses radio telescopes, either interferometers or single dishes, to map large-scale structures using the 21cm emission line, a forbidden hyperfine transition arising from the spin-flip of the electron relative to the proton in a neutral hydrogen (H\,\textsc{i}) atom. Unlike the aforementioned optical spectroscopic and photometric surveys, which painstakingly resolve individual galaxies to map large-scale structures, IM experiments instead measure the integrated emission from unresolved galaxies, combined into a single pixel. Interferometric IM experiments do this using short baselines (low angular resolution) but increased sensitivity and survey speed with many antennas, thereby achieving a high signal-to-noise ratio per pixel \citep{battye_neutral_2004,loeb_possibility_2008,pritchard_evolution_2008,chang_baryon_2008, seo_ground-based_2010}.  

Observational systematics are the primary bottleneck to any 21cm survey that aims to measure the BAO signal. The systematics that 21cm arrays suffer from are both astrophysical and instrumental. Among the instrumental ones, primary beam chromaticity is a prime concern, as it couples the bright foregrounds into the signal \citep{thyagarajan_foregrounds_2015, thyagarajan_effects_2016}. This is the frequency-dependent spatial response of the radiation pattern or, in radio astronomy terminology, the primary beam (PB hereafter) of a single antenna in an array of either dishes or dipoles. This PB is specific to antenna designs, typically a reflector with a feed at its focus for a dish-feed configuration in higher-frequency IM experiments \citep{saliwanchik_mechanical_2021, bandura_packed_2019, vanderlinde_canadian_2019, wu_tianlai_2021} and often a dipole configuration in lower-frequency Epoch of Reionization (EoR) experiments \citep{lonsdale_murchison_2009,haarlem_lofar_2013, aguirre_precision_2014}. The way the PB response manifests in the visibility data can be understood by considering the visibility equation derived from the van Cittert--Zernike theorem \citep{van_cittert_wahrscheinliche_1934, zernike_concept_1938}. The theorem states that the complex visibility (the spatial coherence function) of a distant, incoherent source equals the Fourier transform of its intensity distribution. The complex visibility $V$ of the sky brightness temperature distribution $T$ at frequency $\nu$ measured by a pair of antennas $i$ and $j$ with identical radiation patterns, i.e. power beam ${B}$, is given by, 

\begin{equation}
\label{vis_integral}
        V^{ij} (\nu) = \int {B}(\boldsymbol{\hat\theta},\nu) ~ T(\boldsymbol{\hat\theta},\nu) ~ \exp \left[-2\pi i ~ \boldsymbol{u_{ij}} \cdot\boldsymbol{\hat\theta}\right] ~ d\Omega
\end{equation}

where $\boldsymbol{\hat\theta}$ is the unit vector towards a direction on the sky and $\boldsymbol{u}$ denotes the baseline vector that connects the antenna pair in wavelength ($\lambda$) units. To put it simply, the observed visibilities are PB-weighted sky brightness in Fourier space, making the PB a convolution kernel in the $uv$ plane. Note that the equation has no explicit time dependence. Still, in practice, especially in a drift-scan telescope, this is important, since the sky rotates while the telescope is fixed with respect to the ground, making $T$, and hence $V$, time-dependent.

There are numerous practical challenges in a precision 21cm cosmology experiment aimed at demystifying the nature of dark energy. Among others, these include Radio Frequency Interference (RFI) \citep{offringa_impact_2019}, foregrounds on their own, foregrounds coupled with instrument chromaticity \citep{bernardi_foregrounds_2009, pober_importance_2016}, mutual coupling between the dishes in a compact array \citep{kern_mitigating_2019, josaitis_array_2022}, pointing errors \citep{orosz_mitigating_2019}, and non-redundant effects in a redundant array \citep{dillon_redundant-baseline_2020}. Each of them requires careful characterization and control to be accounted for in the data during analysis. One can categorize the instrumental effects as direction-dependent (DD) and direction-independent (DI). DI effects, such as complex gains, can usually be pulled out of the visibility integral and thus solved for during the calibration stage of the analysis procedure. However, the DD effects are absorbed into the visibilities in a real observation (i.e., sky-integrated), so one has to forward-model these effects to characterize them. To achieve desirable precision in constraining cosmological parameters, these forward models should be highly accurate and regularly updated. The presence of bright foregrounds mixed with these DD systematics poses severe challenges, as mismodeling can bias the signal detection \citep{liu_data_2020}. 

The inherently spectrally smooth foregrounds, coupled with the chromatic response of the interferometer, manifest a wedge-shaped region, leaving a clean signal window in the high-$k_\parallel$ region of the cylindrical power spectrum, and thus can be isolated \citep{datta_bright_2010, parsons_per-baseline_2012, morales_four_2012}. This technique is called foreground avoidance, and many projects currently follow this approach \citep{pober_what_2014, kolopanis_simplified_2019, trott_deep_2020}. But even in this case, the foregrounds leak out of the ideal wedge horizon and contaminate the clean signal window due to instrumental systematics \citep{barry_calibration_2016, ewall-wice_hydrogen_2016}. Ideally, one should be able to clean all foregrounds and recover signal even in the wedge regions to get the best out of the observations and achieve high-SNR signal detection. Conventional cleaning algorithms, such as PCA, can remove these bright foregrounds but either leave residuals or degrade the cosmological signal \citep{alonso_blind_2015, cunnington_foreground_2023}. Since these are mostly blind subtraction techniques, there is always the issue of preserving the signal. Therefore, a robust method should model the physics of the foregrounds and the cosmological signal, as well as the instrument, to cleanly disentangle them and achieve high-SNR signal detection. 

Multiple ML-based frameworks, especially U-Net-architecture-based foreground-cleaning methods, have been proposed in the past \citep{makinen_deep21_2021,shi_21-cm_2024,chen_stability_2024,sabti_generative_2025}. Still, they either operate on idealized image-space maps, neglecting the instrumental effects introduced by map-making, or rely on supervised methods with labels derived from simulations \citep{li_separating_2019}. To our knowledge, there are currently no instrument-agnostic methods that can filter out foregrounds from observed visibilities without requiring an explicit instrument or sky model. In this work, we develop a self-supervised convolutional autoencoder that identifies the instrument state from 21cm drift-scan visibilities without instrument-state labels. We design an encoder-decoder architecture with dual-head MLP blocks for latent encoding, one for the instrument and one for the sky component, where the latent separation occurs. We also include a \textit{sky-only} decoder block that produces ``clean'' \textit{sky-only} visibilities from the sky latent, and an instrument operator block that reconstructs the \textit{simulated} visibilities from the \textit{sky-only} visibilities and the instrument latent. This work involves careful data preparation and physics-based loss engineering. This work also explores the learned latent space to shed light on the so-called ``black box'' nature of ML models. 

This paper is structured as follows. In Sect. \ref{sim_data}, we describe the simulation procedure of the visibility data for a typical compact dish array in drift-scan mode. The convolutional autoencoder network architecture, training procedure, losses, and metrics are discussed in Sect. \ref{methodology}. In Sect. \ref{results}, the test results are discussed with reference to the latent vectors of both the instrument and the sky. We discuss the success and limitations of this method in its current form, and how it could evolve with further refinements in Sect. \ref{discussion}, and provide a summary in Sect. \ref{conclusion}.   
    

\section{Simulation and Data Preparation}
\label{sim_data}

We simulate drift-scan visibilities using the SERVAL package \citep{crichton_2026_serval}, which uses the well-known m-mode formalism \citep{shaw_all-sky_2014, shaw_coaxing_2015} to quickly generate visibilities by precomputing a sky cache and then customizing the PB, or vice versa, as needed. We precompute the sky cache using SERVAL, and the beam model varies for each instrument seed. The beam model is used only in the simulation; the network never has access to it. The instrumental configuration we chose for this work is a compact dish array with a regularly spaced layout, and all the simulation parameters are given in Table~\ref{table1}. Sky maps are in units of brightness temperature (K). Since the visibility integrates the beam-weighted sky over solid angle (Equation~\ref{vis_integral}), we quote visibilities in $\mathrm{mK~sr}$ throughout.

\begin{table}
\begin{tabular}{ l | >{\raggedright\arraybackslash}m{4cm}  }
\hline
\hline
\multicolumn{2}{c}{Data simulation parameters}\\
\hline
Observation mode & Drift-scan \\
Array latitude & $30^\circ$ S \\
Array pointing & Zenith \\
Frequency range & 400 to 800 MHz\\
No. of channels & 128\\
Array layout & Regular 7m spaced $6 \times 6$ grid \\ 
No. of unique baselines & 16 baselines (from $7$ to $39.6$\,m)\\
Sky model & GSM + H\,\textsc{i} (cora package generated)\\
RA range & $0-360^\circ$ (full RA range)\\
RA samples & 1024 \\
Beam model & Circular Gaussian (with spectral ripple)\\
Effective Aperture size ($D_{eff}$) & 4.6m (Physical size  = 6m)\\
\hline
\end{tabular}
\caption{Visibility simulation parameters}
\label{table1}
\end{table}

We chose a smaller number of frequency channels (128, i.e., a channel width of 3.125 MHz) and a set of 16 unique baselines chosen from a coplanar $6 \times 6$ grid of $7$\,m
spacing, spanning lengths from $7.0$ to $39.6$\,m, so that the resulting total visibility cube from the SERVAL simulation is manageable. For a fixed zenith-pointing telescope and fixed declination, the right ascension axis is the same as the time axis, since the RA on the meridian equals the local sidereal time; after the sky makes one complete rotation, we see the same sky strip the next sidereal day. In practice, these drift-scan experiments would need to run for several months to achieve the sensitivity required for cosmological detection. In contrast, we simulate a single sidereal day per instrument state. The sky we chose comprises two components: the H\,\textsc{i} signal generated from a Gaussian random field output of the \texttt{CORA} package \citep{shaw_radiocosmologycora_2020} and the Global Sky Model (GSM16; \citealt{zheng_improved_2017}) of the radio sky with no additional point sources added. We simulate visibilities for these two sky components separately to compare our results with simulated H\,\textsc{i}-only visibilities once we ultimately achieve complete disentanglement and test our signal recovery. These components can be added together during training, since the visibility is linear in $T$ (Equation~\ref{vis_integral}), provided the instrument model is consistent across them. 

We use a Gaussian beam model for simplicity. However, we plan to use more realistic PBs generated with electromagnetic solvers such as CST Studio Suite \footnote{CST Studio Suite: \url{https://www.3ds.com/products-services/simulia/products/cst-studio-suite.}}, in future work. Here, the assumed PB is a 2-dimensional circular Gaussian function centered at the zenith,

\begin{equation}
    B(\theta) \propto ~ \exp\left(- \frac{{\theta}^2}{2\sigma^2}\right).
\end{equation}

Here ${\theta}$ is the angular distance from the beam center, and $\sigma$ is the standard deviation, which is related to the full-width at half-maximum (FWHM) by the relation,

\begin{equation}
     \sigma = \frac{\text{FWHM}}{ 2\sqrt{2\ln 2}}.
\end{equation}

 To simulate higher-order chromaticity in this beam model, we add a term to the conventional theoretical beam width. Therefore, the FWHM of our simulated instrument is given by,

\begin{equation}    
    \text{FWHM} = C ~ \frac{\lambda}{D_{eff}}
\end{equation}

where 

\begin{equation}    
    C = \mathscr{A} ~\sin\left(2\pi \frac{\nu}{\Delta\nu} n\right)+ 1.029 .
\end{equation}

Here $\lambda = c / \nu$  is the wavelength, $D_{eff}$ is the effective aperture diameter, $\Delta\nu$ is the bandwidth, $n$ is the number of ripple periods with $n = 3$, and the resulting ripple period is 133.33 MHz, and 1.029 is the FWHM factor of the Airy pattern of a uniformly illuminated circular aperture. The ripple amplitude

\begin{equation}
    \mathscr{A} \sim \mathcal{N}(0,\,0.1^2)
\end{equation}

is drawn from a normal distribution and is constant over frequency and RA for a given seed. We refer to this quantity as the instrument state throughout the paper, as it is the chromatic systematic we consider here. This assumption of ripple-like chromaticity from the main lobe of the PBs is inspired by real observed phenomena, usually attributed to standing waves between the dish and the feed, as reported in HIRAX simulations \citep{sampath_primary_2026}, CHIME data \citep{collaboration_detection_2026} and MeerKAT simulations \citep{matshawule_h_2021}. 

There are no redundant baselines in this simulation, as we simulate only the 16 unique baselines in the assumed $6 \times 6$ array; with no noise and dish-to-dish variations, redundant baselines would give identical visibilities. We also assume there are no dish-to-dish variations, so the PB is uniform throughout the array. We let the instrument drift through the sky strip centered at $-30^\circ$ declination for 1024 RA samples. The dataset is constructed in the shape of $[Re/Im, N_{s}, N_{r}, N_{b}, N_{\nu}] = [2,100,1024, 16,128]$, where each dimension represents the real or imaginary part, number of instrument seeds, number of RA pointings, number of baselines, and number of frequency channels, respectively. Note that the RA sampling ($\approx 0.35^\circ$) is much finer than the beam FWHM ($\approx 9.6^\circ$ at 400 MHz and $\approx 4.8^\circ$ at 800 MHz), so adjacent RA samples are strongly correlated, and that's why we shuffle the RA indices during training. Although the full RA range was simulated with 1024 RA samples, we train and evaluate on only half of it (512 RA samples) to reduce computational cost. We find this subset sufficient for the network to learn the instrumental response and to generalize across the interleaved sky pointings; however, generalization to an entirely unseen sky is only partial (Appendix~\ref{vis_recon}). Because the RA indices are shuffled before the train/test split, test pointings within this half are interleaved with training pointings and share the same sky structure as their neighbors. The unused half of the sky, therefore, offers a stricter test; it contains sky regions that the network has never seen at any pointing. We use it to assess generalization to an entirely unseen sky and the fidelity of the \textit{reconstructed} visibilities, and present the results in Appendix~\ref{vis_recon}. So the actual shape of the visibility cube used before the data split is $[2, 100, 512, 16, 128]$. 

\section{Methodology}
\label{methodology}

\subsection{Network architecture}
\label{net_arch}

\begin{figure*}
    \centering
    \includegraphics[width=1\linewidth]{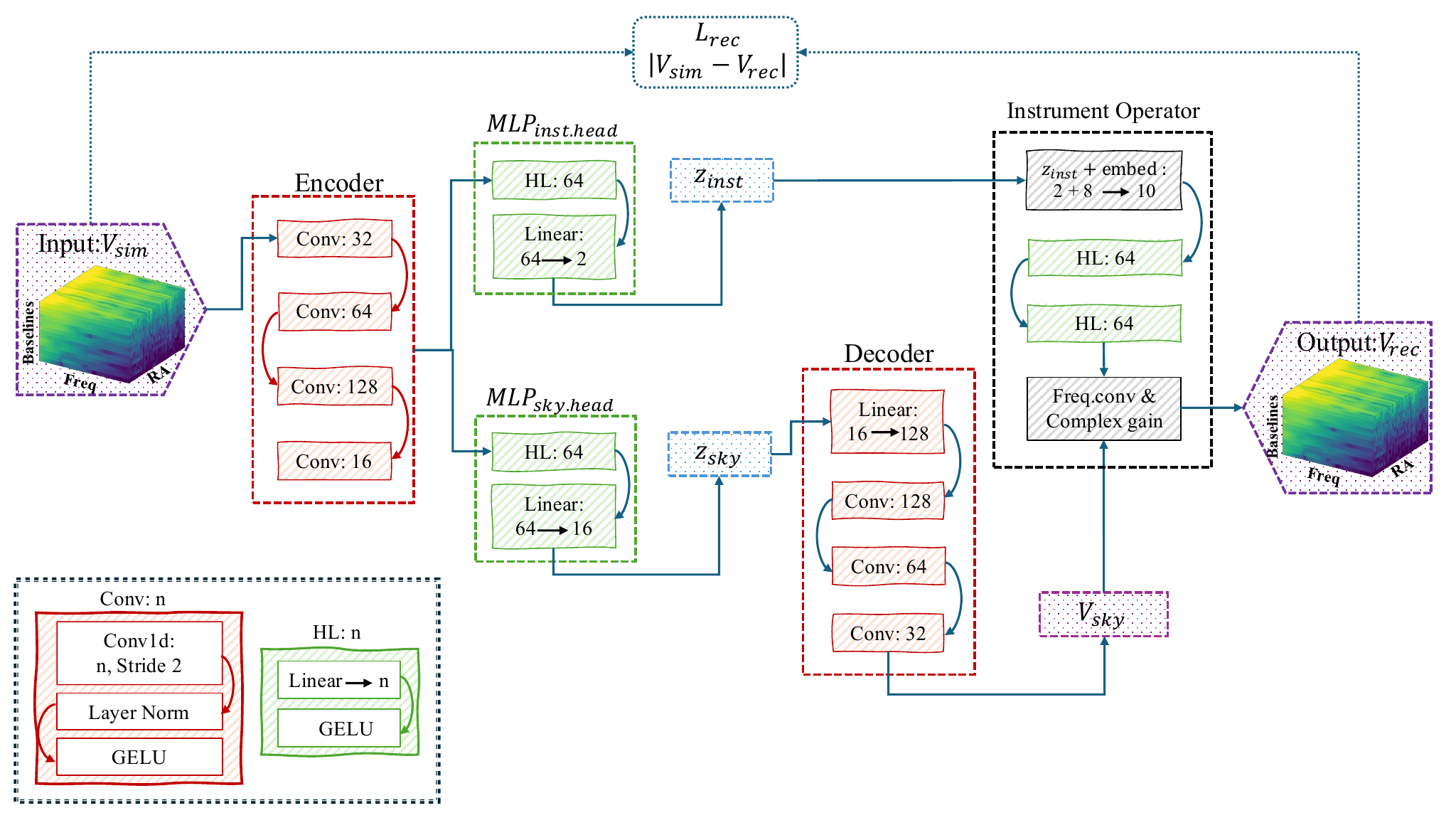}
    \caption{Network architecture. The encoder splits $V_\mathrm{sim}$ into $z_\mathrm{inst}$ and $z_\mathrm{sky}$; the sky-only decoder gives the clean $V_\mathrm{sky}$, and the $\texttt{InstrumentOperator}$ applies $z_\mathrm{inst}$ to $V_\mathrm{sky}$ to give $V_\mathrm{rec}$. The legend for the \texttt{Conv} and \texttt{HL} blocks is in the bottom left. }
    \label{fig:architecture}
\end{figure*}

A typical encoder–decoder network compresses its input into a low-dimensional latent representation and then reconstructs it from that representation \citep{hinton_reducing_2006}. We modify a traditional autoencoder setup by splitting the latent into two blocks, following earlier work that assigns separate latent blocks to different factors of variation \citep{kulkarni_deep_2015, mathieu_disentangling_2016}. Because the network must reconstruct well, the latent is forced to capture the essential structure of the data, and splitting the latent into separate blocks lets each block specialize in a different part of that structure. This suits our problem for two reasons. First, the reconstruction objective needs no instrument-state labels, beam models, or foreground priors: the pairing of observations alone tells the network what is shared between them (the sky for the same RA, the instrument for the same seed) and what differs, similar to how grouped observations are used to separate shared and sample-specific latents in \cite{bouchacourt_multi-level_2018}. This pairing is a weak form of supervision \citep{locatello_weakly-supervised_2020, kugelgen_self-supervised_2022}, but no target values are needed. Second, an encoder alone could learn to detect instrumental changes but would provide no way back to the data, whereas a decoder would make the learned representation usable for cleaning. Our aim in this work is to recover the instrument state from the instrumental latent reliably; full disentanglement of the instrument and sky is left to future work.

Fig.~\ref{fig:architecture} shows the architecture with the shortened \texttt{Conv} and \texttt{HL} (Hidden Layer) blocks with \texttt{GELU} activation \citep{hendrycks_gaussian_2023} shown in distinct colors. A shared encoder maps each input visibility to an embedding, which then feeds into two MLP heads. The instrument head learns a sky-invariant instrumental latent, $z_\mathrm{inst}$, and the sky head learns an instrument-invariant sky latent, $z_\mathrm{sky}$. This separation is encouraged by two consistency losses, described in Sect.~\ref{losses}. Unlike most encoder–decoder networks, ours has only a sky decoder, which maps $z_\mathrm{sky}$ back to a visibility $V_\mathrm{sky}$. This choice makes the model architecture asymmetric, but the sky latent can in principle be used to infer true sky visibilities free of instrumental systematics, which is the ultimate goal of this setup. There is no instrument decoder, and this is a deliberate inductive bias: the instrument beam emits nothing on its own and only distorts the sky it observes. Hence, an instrument-only visibility has no physical meaning. This restricts the model to multiplicative and convolutive systematics; additive systematics such as cross-talk are not captured. Instead, an $\texttt{InstrumentOperator}$ block applies the instrumental effect encoded in $z_\mathrm{inst}$ to $V_\mathrm{sky}$, producing the \textit{reconstructed} visibility $V_\mathrm{rec}$ that is compared with the input. The instrument latent never passes through the decoder; it only sets the operator's coefficients. The operator is therefore linear in the sky visibility, approximating a real instrument response, while the mapping from $z_\mathrm{inst}$ to those coefficients can be nonlinear. It is an approximation because a real beam change acts as a convolution in the $uv$ plane that mixes baselines, which a per-baseline operator cannot represent exactly. Note also that $V_\mathrm{sky}$ is identified only up to an invertible transformation common to all instrument states: any such transformation (e.g., applying the instrument-averaged beam response) can be absorbed into the operator, with its inverse applied to $V_\mathrm{sky}$, leaving $V_\mathrm{rec}$ unchanged. The sky latent therefore recovers the sky as seen through this shared response, not the true sky.

The network’s basic input unit is a block: the \textit{simulated} visibilities $V_\mathrm{sim}$ for a single (seed, RA) pair, where the seed fixes the instrument state and the RA fixes the sky pointing. Each block has shape $[N_b, 2, N_\nu]$, with the real and imaginary parts stacked as two channels. The encoder processes each block independently, so it has no access to neighboring pointings or to the continuity of the sky along the scan; information is combined across blocks only to compute losses in the MLP heads, described below. The encoder is a three-layer 1D convolutional network applied independently to each baseline slice. Its stride-2 convolutions reduce the 128-sample frequency axis by a factor of 8, and a linear projection then maps the result to a 16-dimensional embedding $h$, one per baseline. The sky decoder is the mirror of the encoder: a linear projection followed by a stack of 1D transposed convolutions that restores the full frequency axis.

The two MLP heads pool the embedding along complementary axes. Let $h_{s,r,b}$ denote the encoder embedding for seed $s$, RA $r$, and baseline $b$. The instrument head acts on the baseline-averaged embedding of a single block,
\begin{equation}
    z_\mathrm{inst}^{(s,r)} = \mathrm{MLP}_\mathrm{inst}\!\left( \frac{1}{N_b} \sum_{b=1}^{N_b} h_{s,r,b} \right),
    \label{eq:zinst_block}
\end{equation}
giving one latent per (seed, RA). For reconstruction, these are further averaged over the $N_r$ RAs sharing seed $s$ that are processed together in a batch,
\begin{equation}
    z_\mathrm{inst} \equiv \frac{1}{N_r} \sum_{r=1}^{N_r} z_\mathrm{inst}^{(s,r)},
    \label{eq:zinst_pooled}
\end{equation}
so that only what is common to a single instrument state survives; we drop the superscript $s$ for brevity, and $z_\mathrm{inst} \in \mathbb{R}^2$ without superscripts always refers to this baseline- and RA-averaged latent. The sky head does the converse: it acts on each baseline separately, $z_\mathrm{sky}^{(s,r,b)} = \mathrm{MLP}_\mathrm{sky}(h_{s,r,b})$, and for decoding these are averaged over the $N_s$ seeds sharing RA $r$ in a batch,
\begin{equation}
    z_\mathrm{sky} \equiv \frac{1}{N_s} \sum_{s=1}^{N_s} z_\mathrm{sky}^{(s,r,b)},
    \label{eq:zsky_pooled}
\end{equation}
giving one $z_\mathrm{sky} \in \mathbb{R}^{16}$ per (RA, baseline) pooled per observation during test, in which only what is common to a single sky survives. We keep $z_\mathrm{inst}$ deliberately small because the instrument parameter we want to recover is a single number, $\mathscr{A}$; we use two dimensions since a one-dimensional latent performed worse (Sect.~\ref{discussion}). In principle, the size of each bottleneck controls how much information its latent can hold, and so which components of the visibilities it learns; reworking these dimensions is a possible route to full disentanglement. Finally, the $\texttt{InstrumentOperator}$ uses $z_\mathrm{inst}$ to set the coefficients of a complex gain and a convolution along the frequency axis, which it applies to $V_\mathrm{sky}(r, b, \nu)$ to give $V_\mathrm{rec}(r, b, \nu)$.
For each baseline, a small two-layer MLP takes $z_\mathrm{inst}$, together with a learned embedding of the baseline’s length bin, and outputs the coefficients of two operations applied to $V_\mathrm{sky}$: a frequency-dependent complex gain and a short frequency-domain filter. Baselines in the same length bin therefore share the same response. The gain is a sum of $N_c$ cosine modes in normalized frequency $\tilde{\nu} \in [0, 1]$, and the filter $\kappa_b(m)$ has length $L$ and is added to the identity, so that an untrained operator leaves the sky unchanged:
\begin{equation}
g(b,\nu) = 1 + \sum_{k=0}^{N_c-1} c_{b,k} \cos(\pi k \tilde{\nu}),
\qquad
\kappa_b(m) = \delta_{m,0} + \tilde{\kappa}_{b,m},
\end{equation}
with complex coefficients $c_{b,k}$. The reconstructed visibility is then
\begin{equation}
V_\mathrm{rec}(r,b,\nu) = g(b,\nu) \sum_{m} \kappa_b(m), V_\mathrm{sky}(r,b,\nu-m),
\end{equation}
where $m$ runs over the $L$ channel offsets centred on zero. We use $N_c = 8$, filter length $L = 5$ and $N_\mathrm{bin} = 4$ length bins, so each instrument state is described by just 84 numbers, which set the full complex response across all $N_b = 16$ baselines and $N_\nu = 128$ channels.

In summary, each observation ($V_\mathrm{sim}$) is encoded into an instrumental latent $z_\mathrm{inst}$ and a sky latent $z_\mathrm{sky}$. Decoding $z_\mathrm{sky}$ gives the clean \textit{sky-only} visibility $V_\mathrm{sky}$, and applying the instrument operator to it gives the \textit{reconstructed} visibility $V_\mathrm{rec}$. We use these names throughout the rest of the paper.

\subsection{Loss}
\label{losses}
The total loss computed per training or validation step has three components and is given by
\begin{equation}
    L_{total} =\lambda_{rec} L_{rec} + \lambda_{inst} L_{inst} + \lambda_{sky} L_{sky}  \, .
\end{equation}

Here,
\begin{quote}
     
\begin{enumerate}

\item  $L_{rec}$ is the reconstruction loss that penalizes the difference between the noise-free \textit{simulated} and \textit{reconstructed} visibilities,
\begin{equation}
    L_{\text{rec}} = \frac{1}{N_s N_r N_b N_{\nu}} \sum_{s,r,b,\nu} \left| V_{\mathrm{rec}}^{(s,r,b,\nu)} - V_{\mathrm{sim}}^{(s,r,b,\nu)}\right|^2
\end{equation}

and enforces that the full generative path (sky-decoder + \texttt{InstrumentOperator}) can reproduce the data. Without this loss, the latent could encode anything with no actual grounding in the true signal. 
\\
\item $L_{inst}$ is the instrument consistency loss that penalizes disagreement between $z_\mathrm{inst}$ encoded from two disjoint RA subsets, $\mathcal{R}_1$ and $\mathcal{R}_2$, of the same scan and is given by
\begin{equation}
\label{linsteqn}
    L_{inst} = \frac{1}{N_s} \sum_s \Vert z_\mathrm{inst}^{(s,\mathcal{R}_1)} - z_\mathrm{inst}^{(s,\mathcal{R}_2)} \Vert ^2 .
\end{equation}

This enforces RA-invariance of $z_\mathrm{inst}$, i.e., the instrument state does not change as the sky drifts through the beam. The encoder could put RA-dependent sky information into $z_\mathrm{inst}$ without this term in the loss. 
\\
\item $L_{sky}$ is the sky consistency loss that penalizes disagreement between $z_\mathrm{sky}$ encoded from two disjoint seed subsets, $\mathcal{S}_1$ and $\mathcal{S}_2$, seeing the same RA,
\begin{equation}
\label{lskyeqn}
    L_{sky} = \frac{1}{N_r} \sum_r \Vert z_\mathrm{sky}^{(\mathcal{S}_1,r)} - z_\mathrm{sky}^{(\mathcal{S}_2,r)} \Vert ^2 .
\end{equation}
This is similar to $L_{inst}$ but enforces seed- or instrument-invariance of the sky latent. 

\end{enumerate}
\end{quote}

Here, $\lambda_{rec}$, $\lambda_{inst}$, and $\lambda_{sky}$ are weights that we set to unity in this procedure, giving all losses equal weight in the total loss. All three losses are jointly necessary for this mechanism to succeed, as $L_{rec}$ alone would underdetermine and allow unfaithful reconstruction, including one where both latents encode a mixture of instrument and sky. $L_{inst}$ and $L_{sky}$ reduce this degeneracy by imposing invariance constraints: each latent must be blind to the variation of the other factor. This loss setup enables training without instrument-state labels, and these invariances are derived directly from the physical structure of the observation, making this method self-supervised. For real data, this relies on the instrument remaining stable within a scan and on observing the same sky under different instrument states (e.g., on different nights). 

\subsection{Training procedure}
\label{training}
\begin{figure*}
    \centering
    \includegraphics[width=\linewidth]{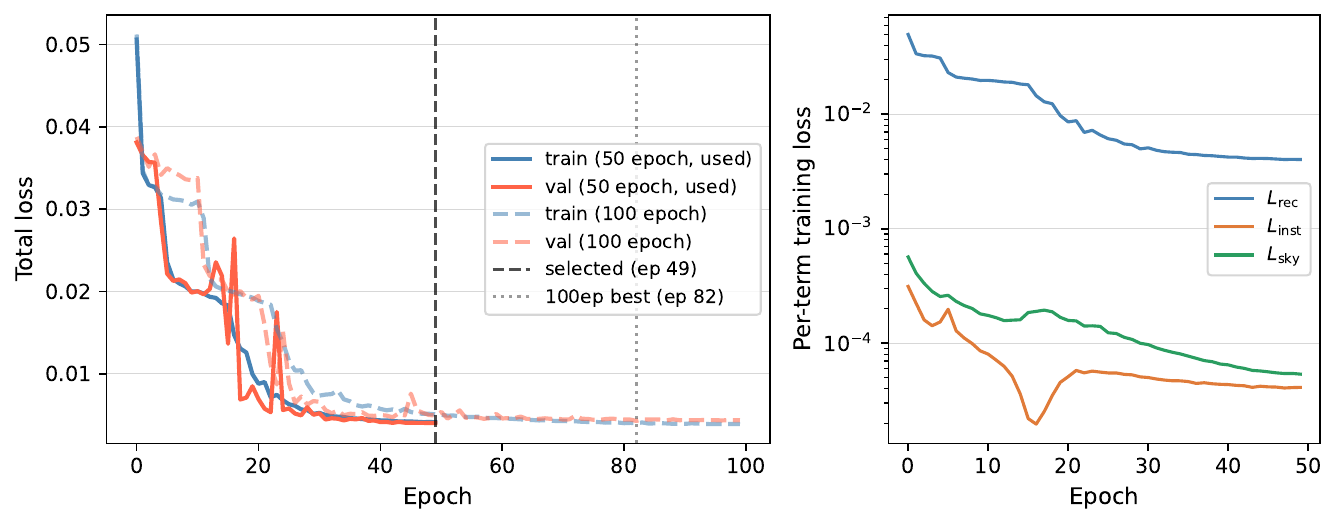}
    \caption{Loss curves for the reported model. Left: total training (blue) and validation (red) loss for the 50-epoch run used in this work (solid) and a 100-epoch run with the same configuration (dashed); vertical lines mark the selected checkpoints. After an initial plateau, the loss drops sharply once the instrument path engages, and the longer run converges to the same level. Right: individual training-loss terms. $L_\mathrm{rec}$ dominates, while $L_\mathrm{inst}$ and $L_\mathrm{sky}$ remain about two orders of magnitude smaller. In both training and validation, inputs include noise augmentation and targets are noise-free.}
    \label{fig:losscurves}
\end{figure*}

The 100 seeds are the instrument states produced by varying $\mathscr{A}$, using seed numbers 0–99 for the normal-distribution draw described in Sect.~\ref{sim_data}. We split the visibility cube 70:15:15 into training, validation, and test sets independently along both the seed and RA axes, giving 70/15/15 seeds out of 100 and 358/76/78 RAs out of 512. Because the two axes are split independently, the test set is held out in both directions: the model never sees those instrument states or those sky positions during training. A model that memorized the instrument states would fail on held-out seeds. However, since adjacent RAs are strongly correlated (Sect.~\ref{sim_data}), held-out RAs are not independent sky, and the unseen half of the sky (Appendix~\ref{vis_recon}) is the stricter test against sky memorization. Because the simulations are noise-free, white Gaussian noise is added to the encoder input during training at a level drawn log-uniformly over $\eta \in [0.1, 10]$ per block, where $\eta$ is the noise RMS in units of the instrument effect itself (the across-seed scatter at fixed RA), with a quarter of blocks left clean and the reconstruction target kept noise-free. However, we do not evaluate the network's robustness to thermal noise, which is left for future work. 

Each training batch contains 64 blocks, built from 8 training seeds $\times$ 8 training RAs, so the encoder output for a batch has shape $[64, N_b, 16]$, one embedding per (seed, RA, baseline). The consistency pairs for the losses are formed on the fly from batch indices rather than precomputed. For $L_\mathrm{inst}$, the 8 RAs sharing a seed $s$ are split into two disjoint halves, $\mathcal{R}_1$ and $\mathcal{R}_2$. Each half is mean-pooled over baselines and passed through the instrument head, giving two latents, $z_\mathrm{inst}^{(s,\mathcal{R}_1)}$ and $z_\mathrm{inst}^{(s,\mathcal{R}_2)}$, that share the instrument but differ in sky (Equation~\ref{linsteqn}). For reconstruction, the instrument-head outputs are averaged over all 8 RAs of seed $s$ in the batch, giving a single pooled $z_\mathrm{inst}^{(s)}$ that the $\texttt{InstrumentOperator}$ applies to every block with that seed. Similarly, for $L_\mathrm{sky}$, the 8 seeds sharing an RA are split into two disjoint halves, $\mathcal{S}_1$ and $\mathcal{S}_2$, passed through the sky head per baseline and mean-pooled across seeds, giving two latents that share the sky but differ in instrument state ($z_\mathrm{sky}^{(\mathcal{S}_1,r)}$ and $ z_\mathrm{sky}^{(\mathcal{S}_2,r)}$ in  Equation~\ref{lskyeqn}). Averaging over all seeds gives the pooled $z_\mathrm{sky}$ used for decoding.

After the forward pass and loss computation for a batch, backpropagation computes the gradient of the total loss with respect to every learnable parameter. We then clip the gradients so that their total norm does not exceed 5.0, rescaling them as needed. Without clipping, a spike in one loss term (such as the one near epoch 20 in Fig.~\ref{fig:losscurves}) could produce an excessively large update and destabilize training; clipping limits the step size while preserving the gradient direction. The \texttt{Adam} optimizer \citep{kingma_adam_2017} (with $L_2$ weight decay $10^{-5}$ applied to the gradient) then updates all parameters in a single step. At the end of each epoch, a \texttt{CosineAnnealingLR} \citep{loshchilov_sgdr_2017} scheduler ($T_\mathrm{max} = 50$ epochs, $\eta_\mathrm{min} = 0$) adjusts the learning rate, annealing it from $10^{-3}$ towards zero over the full run.

Fig.~\ref{fig:losscurves} shows the training and validation loss curves (left) and the contribution of each loss term during training (right). The training and validation curves track closely, so the model shows no sign of overfitting, even without regularization beyond weight decay and noise augmentation. There are occasional spikes, but the model recovers from each and training stabilizes. The per-term loss curves show a clear separation in scale: $L_\mathrm{rec}$ is about two orders of magnitude larger than the consistency terms, as expected, since the visibilities and the low-dimensional latent vectors have very different scales. The consistency losses, therefore, do not dominate the total loss under our choice of equal weights, $\lambda = 1$. We show curves for both a 50-epoch run (solid) and a 100-epoch run (dashed). Both converge, and the longer run reaches a lower validation loss, but it also yields a lower linear-probe $R^2$ for recovering $\mathscr{A}$ (Sect.~\ref{results}). We therefore use the 50-epoch checkpoint. This suggests that, with longer training, the consistency terms reorganize the latent space in ways that make it harder to decode linearly. The loss curves confirm that optimization succeeded; the evaluation metrics in Sect.~\ref{results} establish what the learned representations encode.

\subsection{Evaluation}
\label{metrics}

To evaluate our model, we use two standard metrics and define two model-specific diagnostics based on them.

\subsubsection{Metrics}
\begin{quote}
\begin{enumerate}

\item \textbf{NMSE:} A normalized mean squared error to evaluate the reconstruction of the visibilities given by
\begin{equation}
    NMSE = \frac{ \sum_i \left| V_i - \hat{V_i} \right| ^2}{\sum_i \left| V_i \right| ^2}
\end{equation}
where $V_i$ and $\hat{V}_i$ are the \textit{simulated} and \textit{reconstructed} visibilities and $i$ runs over all (seed, RA, baseline, frequency) elements of the evaluated set.

\item \textbf{$R^2$ linear probe:} During testing, a linear probe \citep{alain_understanding_2018} between true instrument state $\mathscr{A}$ and instrument-latent $z_\mathrm{inst}$, fitted on the training seeds, is evaluated on the test seeds using the $R^2$ score,
\begin{equation}
    R^2 = 1 - \frac{\sum_{s=1}^{N_{test}} (\mathscr{A}_s - \hat{\mathscr{A}}_s)^2}{\sum_{s=1}^{N_{test}} (\mathscr{A}_s - \bar{\mathscr{A}})^2}
\end{equation}

where $\hat{\mathscr{A}}$ is the true instrument state, $\bar{\mathscr{A}}$ is the mean of the true amplitudes over the test seeds and  ${\mathscr{A}}_s $ is the recovered instrument state using a linear regression of the form, 
\begin{equation}
    {\mathscr{A}}_s = \mathrm{w}^\top {z}_{inst} + \beta .
\end{equation}

Here, w and $\beta$ are the slope and intercept of the linear probe regression. 

\end{enumerate}
\end{quote}

\subsubsection{Diagnostics}

\begin{quote}
\begin{enumerate}
\item \textbf{Cross-probe matrix:} This diagnostic is a separation test, with which one can find how much the latents are separated from each other. To measure that, we use the above-mentioned $R^2$ linear probe, but cross-regress the true instrument state $\mathscr{A}$ and a sky proxy ${S}$ with $z_\mathrm{inst}$ and $z_\mathrm{sky}$. ${S}$ is the logarithm of the total power, a scalar stand-in for sky structure; it is only a coarse summary of the sky at that RA pointing. This gives four components that form the cross-probe matrix $\mathbf{M}$ given by 
\begin{equation}
    \mathbf{M} = \begin{bmatrix} R^2 (z_\mathrm{inst} \rightarrow \mathscr{A}) & R^2 (z_\mathrm{inst} \rightarrow {S}) \\ R^2 (z_\mathrm{sky} \rightarrow \mathscr{A}) & R^2 (z_\mathrm{sky} \rightarrow {S}) \end{bmatrix}
\end{equation}

\item \textbf{Swap test:} This diagnostic probes the sky latent by substitution: holding $z_\mathrm{inst}^{(s)}$ and the target visibility $V^{(s,r)}$ fixed, we decode with a $z_\mathrm{sky}$ (unpooled per-observation) taken from a different seed or a different RA and compute the reconstruction NMSE, which we denote $\varepsilon$,
\begin{equation}    
\varepsilon(\tilde{s},\tilde{r}) \;=\;
\mathrm{NMSE}\Bigl({V}\bigl(z_\text{inst}^{\,s},\,z_\text{sky}^{\,\tilde{s},\tilde{r}}\bigr),\;V^{\,s,r}\Bigr)
\end{equation}

with $\varepsilon_\mathrm{self}$ computed with the unchanged latents, which is the true reconstruction NMSE for that test set, 
\begin{equation}
\varepsilon_\mathrm{self} = \varepsilon(s,r) .
\end{equation}

The value encoded from a different seed at the same RA pointing is
$\varepsilon_\mathrm{swap\text{-}seed}$, 
\begin{equation}
\varepsilon_\mathrm{swap\text{-}seed} = \varepsilon(s',r) ,
\end{equation}
and the value encoded from the same seed but at different RA pointings is $\varepsilon_\mathrm{swap\text{-}RA}$, 
\begin{equation}
\varepsilon_\mathrm{swap\text{-}RA} = \varepsilon(s,r') .
\end{equation}

\end{enumerate}
\end{quote}

\section{Results}
\label{results}
\begin{figure*}
    \centering
    \includegraphics[width=\linewidth]{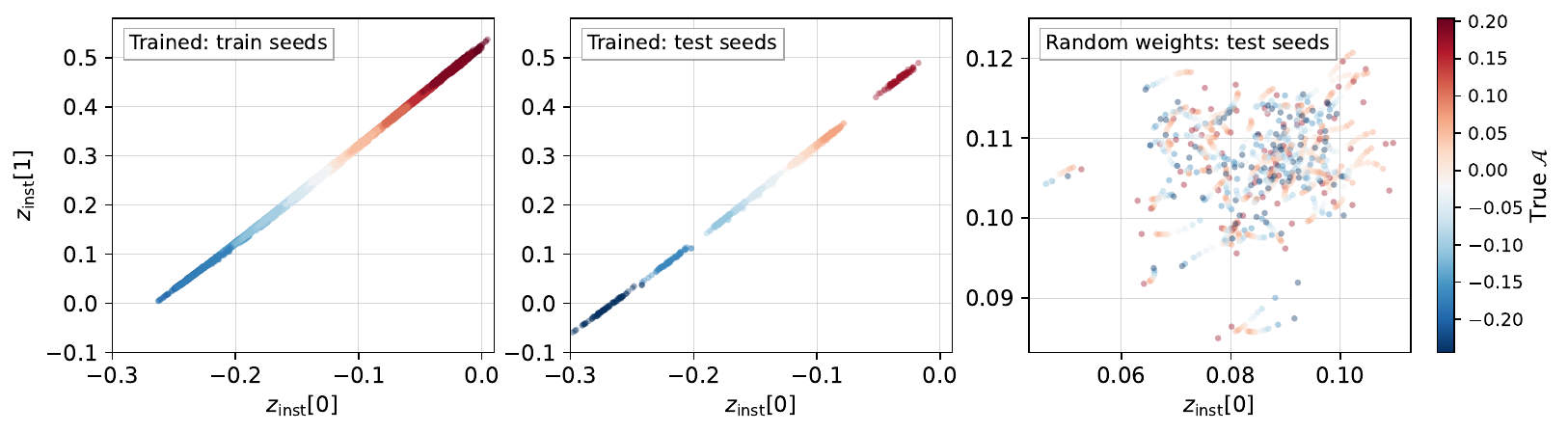}
    \caption{Per-pointing instrument latents $z_\mathrm{inst}^{(s,r)}$, coloured by the true $\mathscr{A}$. \textbf{Left:} training seeds at training RAs. \textbf {Center :} held-out seeds at held-out RAs --- the seed and RA splits are independent, so neither axis was seen in training. \textbf{Right:} the same held-out data encoded with untrained weights. The components lie along a curve in the first two panels, with $\mathscr{A}$ readily identified, and the ordering survives when both axes are held out; the random encoder gives scattered components confined to a much smaller range (note the axis limits). Comparing the first two panels indicates generalization; the last two indicate the effect of training. }
    \label{fig:lat_inst}
\end{figure*}

The objective of the study is not visibility reconstruction, but rather to recover the unseen instrument state $\mathscr{A}$ from the trained network using test seed datasets. We note, however, that the accuracy of visibility reconstruction is vital for the disentanglement mechanism to work, and we discuss this in Appendix \ref{vis_recon}. In this section, we focus on recovering $\mathscr{A}$ from $z_\mathrm{inst}$ and evaluating the network's ability to disentangle the sky and the instrument, each with its own latent. The two instrument-latent vector components align along a curve, and the true $\mathscr{A}$, color-coded on top of it, shows that the encoding is monotonic, as shown in Fig. \ref{fig:lat_inst}. In this figure, we also show a random encoder baseline for comparison, which is the same network but with no trained weights, using the frozen random initialization \footnote{This is a common practice in ML architectures, especially self-supervised methods, where it serves as a control and performance baseline for assessing how much of the downstream task's success stems from the raw architectural geometry versus the learned weights \citep{pilault_impressive_2020, seo_state_2021, bhattacharjea_encoder-decoder_2023}.}. The trained seeds established that the structure exists and that they lie along a monotonic path in $\mathscr{A}$ rather than being scattered. The same trend follows in the held-out test seeds, establishing that the network did not memorize but truly generalized. The random encoder showing no structure confirms that the network's behavior is learned rather than a property of the architecture or the input statistics.

\begin{figure*}
    \centering
    \includegraphics[width=0.9\linewidth]{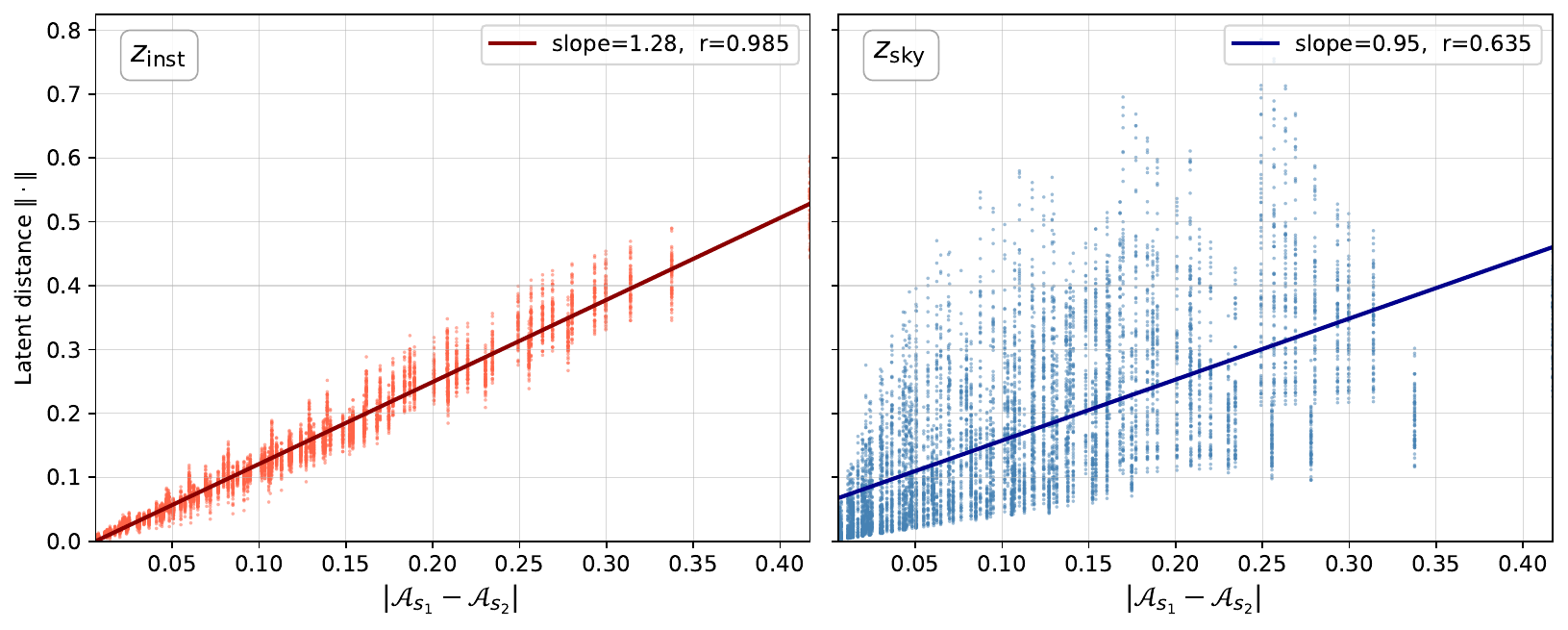}
    \caption{Latent sensitivity to the true instrument-state difference. Euclidean distance between the latent vectors of two seeds, as a function of their instrument-state difference $|\Delta\mathscr{A}|$, for $z_\mathrm{inst}$ and $z_\mathrm{sky}$. Lines show ordinary least-squares fits, with the slope and Pearson $r$ for each latent with respect to $|\Delta\mathscr{A}|$.} 
    \label{fig:lat_sens}
\end{figure*}

Before using the defined $R^2$ linear probe, we can assess whether the latent tracks the true instrument state by using a classical Euclidean distance measure. Fig. \ref{fig:lat_sens} shows the sensitivity of the latents with respect to the true $\mathscr{A}$. From the left panel of the figure, we can infer that even without a linear probe, the $z_\mathrm{inst}$ latent distance scales with instrument difference ($\Delta \mathscr{A} = |\mathscr{A}_{s_1} - \mathscr{A}_{s_2}|$) quite clearly. Both panels use the per-$(s,r)$ latents rather than the pooled ones: $z_\mathrm{inst}^{(s,r)}$ before the average over pointings, and $z_\mathrm{sky}^{(s,r)}$ flattened over baselines. Distances are taken over all pairs of test seeds at a \emph{common} RA ($105$ seed pairs $\times$ $78$ pointings $= 8190$ points), so within each pair the sky is identical, and only the instrument state differs. Ordinary least squares yields a slope of $1.28$ and $r = 0.985$. At the same time, the $z_\mathrm{sky}$ distance tracks $\Delta \mathscr{A}$ more weakly, as one would expect. However, the unconfined spread and the $r$-value of 0.635 from the OLS regression indicate that the sky latent still tracks $\mathscr{A}$ moderately, emphasizing that it is not purely \textit{sky-only} either. The slopes are not comparable across panels, as the two latent variables differ in dimension and scale; the contrast lies in the $r$-value.

\begin{figure}
    \centering
    \includegraphics[width=\linewidth]{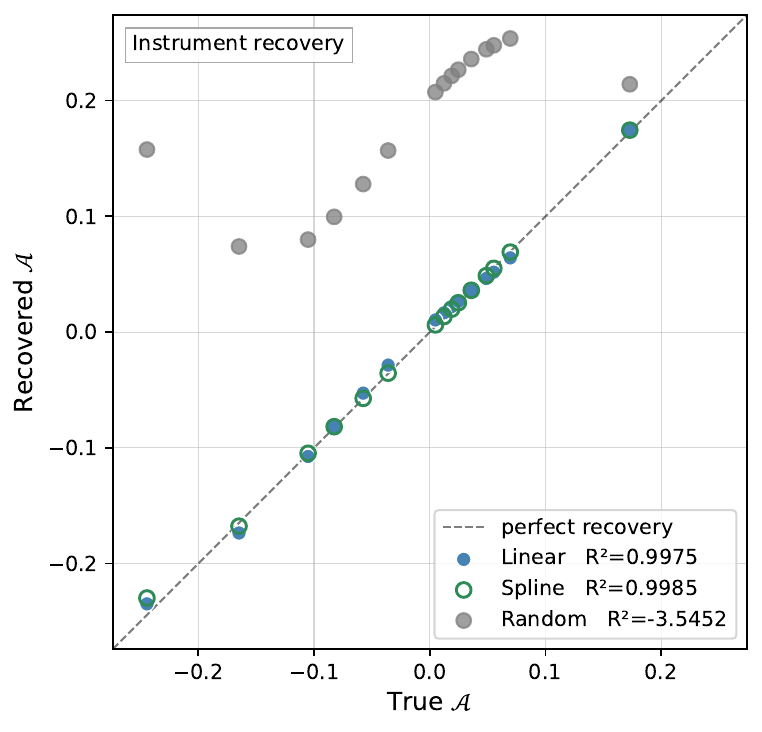}
    \caption{Predicted versus true instrument parameter $\mathscr{A}$ for the 15 held-out test instrument states, each predicted from $z_\mathrm{inst}$ averaged over all held-out RA pointings. Probes are fitted on training simulations only, with linear probe $R^2 = 0.9975$ (blue solid circles) or spline $R^2 = 0.9985$ (hollow green circles). Grey points correspond to the random encoder baseline, which predicts worse than the mean ($R^2 = -3.5452$). The dashed line is perfect recovery.}
    \label{fig:lat_recov}
\end{figure}

Using a linear $R^2$ probe on the $z_\mathrm{inst}$ as defined in Sect. \ref{metrics}, we recover the instrument state $\mathscr{A}$, with an $R^2$ value of 0.9975 for the 15 held-out seeds. A supplementary spline probe predicts $\mathscr{A}$ with an $R^2$ value of 0.9985 for the same test seeds evaluated on the same trained encoder, as we can see in Fig. \ref{fig:lat_recov}, with the dashed line showing the perfect recovery. The spline probe matches the linear probe to within 0.001 in $R^2$. The random encoder representation, when probed with its own linear probe, fails to predict the instrument state with $R^2 = -3.5452$, again demonstrating that the recovery is learned and not arising from the architecture's geometry. Such a strongly negative value means the predictions are worse than simply guessing the mean. Since an untrained encoder’s latents carry no instrument information, they vary mainly with the sky, and the test latents, averaged over held-out pointings, are therefore distributed differently from the training latents on which the calibration was fitted.

We also employ two separation-evaluation diagnostics, as defined in subsection \ref{metrics}. These are all evaluated on the held-out scans: the $15$ test seeds observed at the $78$ test RAs, the seed and RA splits being drawn independently to assess how well the instrument and sky latents are separated. The cross-probe matrix \textbf{M} provides $R^2$ from four independent linear probes, with diagonal elements assessing whether each latent encodes its own factor, and off-diagonal elements assessing whether it leaks the other's factor. Each entry is the $R^2$ of a linear probe, evaluated with grouped 5-fold cross-validation (\texttt{GroupKFold}, $k = 5$, in \texttt{Scikit-Learn}; \citealt{scikit-learn}). All observations sharing a target value are assigned to the same fold, by seed for $\mathscr{A}$ and by pointing for $S$, so no instrument state or pointing appears in both the fitting and scoring folds, and a probe cannot recover a label from other members of its own group. Uncertainties are the standard deviation over 1000 resamplings of the instrument states or pointings. The ultimate aim is to make the diagonal elements high (the sky one cannot reach unity, since $S$ is a coarse proxy) and the off-diagonal elements consistent with zero. 

\begin{figure}
    \centering
    \includegraphics[width=\linewidth]{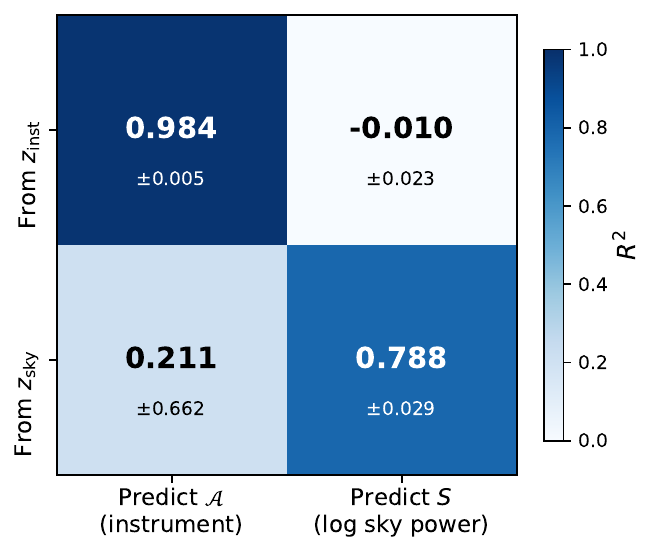}
    \caption{ Cross-probe matrix $\mathbf{M}$: linear-probe $R^2$ for recovering the instrument parameter $\mathscr{A}$ and the sky proxy $S$ from each latent. The top row shows that $z_\mathrm{inst}$ encodes $\mathscr{A}$ well while carrying no detectable sky information; the bottom row hints that some instrumental information remains in $z_\mathrm{sky}$, though with a large uncertainty. Scores are obtained using grouped 5-fold cross-validation, and uncertainties are standard deviations over 1000 resamplings.}
    \label{fig:probe_matrix}
\end{figure}

From Fig. \ref{fig:probe_matrix}, we have the first element showing that the instrument state can be recovered using $z_\mathrm{inst}$ with an $R^2$ value of 0.984. One could argue that this value is lower than the originally reported $R^2$ prediction from the linear probe. The reason is that these cells are computed from unpooled per-(seed, RA) representations, for which all four elements of the matrix are defined, and the probes are cross-validated within the test set. The sky diagonal cell ($R^2 = 0.788$ ) sits below the instrument one, since $\mathscr{A}$ is exactly the parameter that generates the simulated instrumental variation. At the same time, ${S}$ is a coarse summary statistic we came up with. The off-diagonal cells in the matrix show that we can only weakly predict $\mathscr{A}$ from $z_\mathrm{sky}$ with a linear probe ($R^2 = 0.211$), and cannot predict the sky content ${S}$ from $z_\mathrm{inst}$ ($R^2 = -0.01$, consistent with zero). Since a two-dimensional latent could easily encode a scalar such as $S$, the latter is due to the instrument-consistency loss rather than to the low dimensionality of $ z_\mathrm {inst} $. The former, however, shows a positive correlation, hinting that the sky-latent is not free of instrument effects. But the annotated uncertainty on this entry is very wide ($\pm 0.662$),
larger than the estimate itself, unlike the other three entries, which are tighter, with the top-left cell being the tightest. So a further test is needed to establish whether the instrument is still leaking into the sky-latent. 

\begin{figure}
    \centering
    \includegraphics[width=\linewidth]{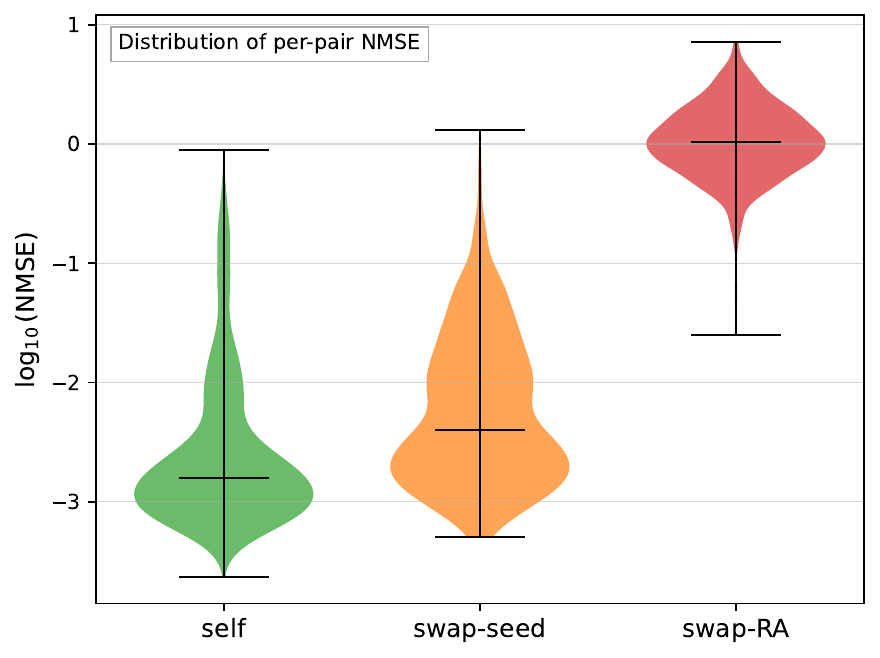}
    \caption{Reconstruction NMSE (log scale) for test observations decoded with their own latents (self, green), or with $z_\mathrm{sky}$ swapped from another instrument state (swap-seed, orange) or another pointing (swap-RA, red). Swapping the instrument state raises the median error by $\approx 2.5\times$, while swapping the pointing raises it by $\approx 660\times$: $z_\mathrm{sky}$ mainly encodes the sky, but still carries some instrumental information.}
    \label{fig:swap}
\end{figure}

Since the probe matrix cannot definitively determine whether the sky-latent carries instrumental information, we perform the swap-test diagnostic on this sky-latent alone. We keep $z_\mathrm{inst}$ unchanged during swapping and use decoding with the swapped seeds and RA pointings. A self component ($\varepsilon_\mathrm{self}$) is kept for reference. The $\varepsilon_\mathrm{swap\text{-}seed}$ and $\varepsilon_\mathrm{swap\text{-}RA}$ are computed for all test seeds, and the log-NMSE distribution for each of these decoded reconstructions is shown in a violin plot (Fig. \ref{fig:swap}). In $\varepsilon_\mathrm{self}$, with both latents taken from the observation itself, the model reconstructs held-out data with a median NMSE of $\approx 10^{-2.8}$, which is low. Still, the extrema are quite wide, indicating room for improvement in reconstruction. In the $\varepsilon_\mathrm{swap\text{-}seed}$ distribution, the sky is identical, so an ideal instrument-invariant $z_\mathrm{sky}$ would leave the error untouched, matching $\varepsilon_\mathrm{self}$. 

The distribution shifts above $\varepsilon_\mathrm{self}$, with the ratio of the medians of these two distributions being $\approx 2.5$, so $z_\mathrm{sky}$ is only partially instrument-invariant, and the gap above self measures the reconstruction penalty from the instrument information it still carries. This acts as the invariance test that the cross-probe matrix could not settle, given its wide spread. When the pointing is swapped instead for the sky-latent to estimate $\varepsilon_\mathrm{swap\text{-}RA}$, this substitution drives the error to $\approx 1$, equaling signal power and comparable to predicting zero. This is a useful validation check, as the reconstruction with RA-swapped $z_\mathrm{sky}$ collapses, confirming that the latent encodes RA-specific features. However, because the target sky differs, this shows that $z_\mathrm{sky}$ is used rather than being instrument-free. The shape of the distribution is also quite compact, showing a Gaussian-like curve. We find that the median ratio of swap-RA to self is $\approx 660$.           

\section{Discussion and future directions}
\label{discussion}
Having been trained on a single-night drift-scan simulation (half of the RA range, 512 RAs used, equivalent to one night's scan) for each of 70 instrument states, without instrument-state labels or beam models (the beam model enters only the simulation), the model recovers the instrument state almost perfectly. Its instrument latent tracks the beam perturbation almost exactly, gives the same answer regardless of which patch of sky is being observed, and carries no detectable sky information as measured by the total-power proxy. This holds for instrument states the network never saw during training and for held-out sky pointings, although the latter are correlated with the training pointings. The main open question is the opposite: how much instrumental signal still leaks into the sky latent, which must be small before the clean visibilities can be trusted for cosmology.

A few caveats are worth mentioning here. After fixing the network design, we performed a hyperparameter search as a sensitivity check, focusing on the consistency-loss weights and the latent size. Across the range tested ($\lambda_\mathrm{inst} \in [0.5, 5]$, $\lambda_\mathrm{sky} \in [0, 5]$; 27 of 29 trained models, the other two setting $\lambda_\mathrm{inst} = 0$ only at $\dim z_\mathrm{inst} = 1$ and so confounding the two), the linear-probe $R^2$ for $\mathscr{A}$ ($0.990$--$0.9998$) varied by no more than the spread between repeated trainings of a single configuration ($R^2 = 0.990$--$0.9995$). The dimension of $z_\mathrm{inst}$ mattered more than the loss weights: a one-dimensional latent reduced $R^2$ to $\approx 0.95$, with or without the instrument-consistency term. Two- and four-dimensional latents performed comparably: the four-dimensional latent was marginally better under a linear calibration ($\Delta R^2 \approx 0.002$), but the difference vanished under a spline calibration. We therefore keep the two-dimensional instrument latent, the smallest one that performs well, since a tighter bottleneck leaves less capacity for sky information to leak in \citep{tishby_information_2000}. We verify that it shows no detectable sky information in Sect. \ref{results} and we discuss hyperparameter search and choice in Appendix \ref{hyper}. We did not isolate $\lambda_\mathrm{inst} = 0$ at $\dim z_\mathrm{inst} = 2$. Its role is nevertheless visible in a full-track run (a complete training at the adopted configuration) with the term removed, rather than a short search trial, where the estimates of $z_\mathrm{inst}$ from disjoint halves of the same scan disagreed by more than the spread between instrument
states, i.e.\ the latent lost the RA invariance that the instrument-consistency term is designed to enforce. 

The current limitations of this framework are the use of controlled simulated data, a single chromaticity nuisance parameter ($\mathscr{A}$), and the assumption that the PB is stable across a scan. We do not evaluate robustness to thermal noise or to RFI flagging here \citep{offringa_impact_2019}; both are left to future work. We acknowledge that this work needs further refinement in its current form, as the disentanglement is not achieved; only the instrument state is recovered. Also, the reconstruction is not perfect, or at least not at the level we need: the H\,\textsc{i} signal is roughly $10^4$--$10^5$ times fainter than the foregrounds, so residuals must eventually be far below the current NMSE of $\sim 10^{-3}$, and the decoded \textit{sky-only} visibility part is not yet comparable to our actual sky. A more scientific metric for assessing the quality of the ``clean'' sky visibility, such as a delay power spectrum \citep{parsons_per-baseline_2012}, would be very useful for comparing it in a real cosmological sense; we defer this to planned future work once we achieve complete disentanglement. Also, if we want to incorporate complex systematics or multiple simultaneous ones into the data, a single instrument latent $z_\mathrm{inst}$ with 2 dimensions will not be sufficient. So, in that case, we would employ multiple instrumental latents, each with its own consistency loss, and the cross-probe matrix would become higher-dimensional; diagonalizing it would be our objective. 

Building on this foundation, we intend to address these limitations in future work. We will first develop a concrete framework to clean up the mixed modes and extract well-behaved sky signals, with an emphasis on preserving the signal. We also plan to optimize this technique, introduce multiple systematics into the data (this work employed only a single systematic), and extend the network’s capabilities to data with RFI flags and realistic thermal noise, moving towards a more comprehensive disentanglement. For real data, the sky-consistency loss will require the same sky to be observed under different instrument states (e.g., from night-to-night variations of the instrument \citep{abdurashidova_first_2022}), and converting $z_\mathrm{inst}$ into a physical parameter will require a calibration. Finally, as the new IM surveys (such as HIRAX \citep{crichton_hydrogen_2022} and CHORD \citep{vanderlinde_canadian_2019}) come online, this framework, in its complete form, will be rigorously tested on real data.

\section{Conclusions}
\label{conclusion}

Instrumental systematics are one of the main concerns of the 21cm Cosmology community, and a framework such as this, which directly addresses them without instrument-state labels or forward models, is expected to be highly valuable once it is validated with real data and made deployable. This marks the initial success in recovering the instrument state, a critical step towards disentangling instrumental and sky signal, providing a proof of concept for a novel approach. The framework presented here is the first step towards an ML-based systematics mitigation and foreground-cleaning pipeline that leverages self-supervised consistency losses \citep{balestriero_cookbook_2023} rooted in physical constraints. This work was started with the ultimate goal of disentangling the sky (foreground and H\,\textsc{i}) and instrumental systematics. The novelty of this method lies in applying consistency losses that enforce self-supervision using realistic physical constraints to drift-scan visibilities, i.e., each latent is invariant to changes in the other factor. 

In this paper, we propose a self-supervised learning approach to disentangle instrumental systematics in 21cm intensity mapping visibilities and present a framework that identifies a single instrumental systematic (i.e., chromaticity). This method requires no model or labels during training; the instrument state is then recovered from the latent with a linear probe calibrated on the training simulations. The ultimate goal, recovering a sky signal entirely free of instrumental effects so that the cosmological signal can be detected, is the natural next step. The sky latent still carries some instrumental imprint, and removing it is the focus of our future work. Nevertheless, this initial result is a proof of concept demonstrating that this approach is viable and shows instrument separation from the sky in the latent space with modest data (70 instrument states $\times$ 358 pointings) and training (50 epochs). Although the original idea is to disentangle the instrumental systematics and, with the ``clean'' \textit{sky-only} visibilities, apply traditional foreground filtering methods \citep{ewall-wice_dayenu_2020,carucci_recovery_2020, kern_gaussian_2021,  spinelli_skao_2022} to extract the H\,\textsc{i} signal, one could extend this framework to a potential sky-mode separation too - separating the foreground and H\,\textsc{i} signal components from visibilities.

\begin{acknowledgements}
      This research was funded by the Swiss National Science Foundation (SNSF) (Grant No: 10001432). For the purpose of open access, a CC BY public copyright licence is applied to any author accepted manuscript (AAM) version arising from this submission. Part of the computations in this work were performed on the Euler cluster operated by the High Performance Computing group at ETH Zürich. We also acknowledge the use of the ilifu cloud computing facility – \url{https://www.ilifu.ac.za}, a partnership between the University of Cape Town, the University of the Western Cape, Stellenbosch University, Sol Plaatje University and the Cape Peninsula University of Technology.
      
      The author would like to thank Martin Kunz for fruitful discussions and mentorship, and Devin Crichton for making their code open-source and providing guidance on simulations. The author would also like to thank Sowmiya Balan for their insightful comments on the paper. 
\end{acknowledgements}

%

\bibliography{refs}
\bibliographystyle{aa}


\begin{appendix}
\section{Hyperparameters}
\label{hyper}
The network design and all hyperparameters in Table~\ref{tab:hyperparameters} were fixed a priori. Afterward, we ran a hyperparameter search with \texttt{Optuna} \citep{akiba_Optuna_2019} as a sensitivity check, not to select the configuration. We varied only three parameters: the two consistency-loss weights, $\lambda_\mathrm{inst}, \lambda_\mathrm{sky} \in \{0, 0.5, 1, 2, 5\}$, and the dimension of $z_\mathrm{inst} \in \{1, 2, 4\}$ (29 trials of 30 epochs each, the sampler visiting 8 of the 75 available configurations). Two trials set $\lambda_\mathrm{inst} = 0$, both at $\dim z_\mathrm{inst} = 1$; the 27 discussed in Sect.~\ref{discussion} are the remainder. The remaining parameters were deliberately not searched: those of the instrument operator limit its capacity so that it cannot reproduce arbitrary visibility structure and absorb sky content into $z_\mathrm{inst}$. Choosing them to maximize a probe score would make the operator as strong as the metric rewards rather than as weak as the physics requires.

The search objective was the linear-probe $R^2$ for $\mathscr{A}$, evaluated on the test instrument states over the whole held-out scan; it was used only to measure sensitivity, never to choose the reported configuration. The training length was likewise fixed in advance at 50 epochs: the 100-epoch run shown in Fig.~\ref{fig:losscurves} is a convergence check, not a candidate selected against the baseline. For every configuration with $\dim z_\mathrm{inst} \geq 2$ the probe gave $R^2 = 0.990$--$0.9998$. Eleven trials that repeated a single configuration spanned $0.990$--$0.9995$, so differences between configurations are comparable to the run-to-run spread of training. A one-dimensional $z_\mathrm{inst}$ reduced $R^2$ to $\approx 0.95$, whether or not the instrument-consistency term was used.

The highest-scoring configuration ($\lambda_\mathrm{inst} = 1$, $\lambda_\mathrm{sky} = 0.5$, $\dim z_\mathrm{inst} = 4$) differs from the reported one ($1$, $1$, $2$). A full-length run with it improved the probe only slightly ($R^2 = 0.9996$ versus $0.9975$ with a linear calibration); the difference vanished with a spline calibration ($ R^2 = 0.9999$ versus $0.9985$). We kept the two-dimensional latent, the smallest one that performs well, because it is replicated across independent training runs and its geometry can be shown directly (Fig.~\ref{fig:lat_inst}).

We did not use the validation loss as the search objective. That loss contains the
weighted consistency terms, so its value changes with $\lambda_\mathrm{inst}$ and
$\lambda_\mathrm{sky}$ themselves and cannot fairly compare configurations. In the sweep, it would have favored $\lambda_\mathrm{inst} = 0$, $\lambda_\mathrm{sky} = 2$, $\dim z_\mathrm{inst} = 1$, the configuration with the poorest detection ($R^2 \approx 0.95$) and the one that drops the instrument-consistency term altogether.

\begin{table}[ht]
\centering
\small
\begin{tabular}{@{} l l @{}}
\toprule
\textbf{Hyperparameter} & \textbf{Value} \\
\midrule
\multicolumn{2}{@{}l}{\textit{Architecture}} \\
\quad Encoder channels   & $[32, 64, 128]$ \\
\quad Decoder channels   & $[128, 64, 32]$ \\
\quad Normalization      & \texttt{LayerNorm} \\
\quad Activation         & \texttt{GELU} \\
\quad Conv kernel size   & 5 \\
\quad Stride             & 2 \\
\quad $z_{\text{inst}}$ dimension & 2 \\
\quad $z_{\text{sky}}$ dimension  & 16 \\
\quad MLP heads & 1 hidden layer, width 64, \texttt{GELU} \\
\quad Instrument operator & 5-tap frequency filter; \\ 
\quad  & 8 cosine gain modes ($16$ coefficients, Re and Im); \\
\quad & 4 baseline-length bins \\
\addlinespace
\multicolumn{2}{@{}l}{\textit{Training}} \\
\quad Optimizer          & \texttt{Adam} \\
\quad Weight decay       & $10^{-5}$ \\
\quad Learning rate      & $10^{-3}$ \\
\quad LR scheduler       & \texttt{CosineAnnealingLR} \\
\quad Epochs             & 50 \\
\quad Batch size         & 64 (8 seeds $\times$ 8 RAs) \\
\quad Gradient-clip norm & 5.0 \\
\quad Noise augmentation & $\eta \in [0.1, 10]$ log-uniform per block \\
\quad Loss weights $\lambda$ & 1.0 \\
\addlinespace
\bottomrule
\end{tabular}
\caption{The a priori hyperparameter choices.}
\label{tab:hyperparameters}
\end{table}

\section{Visibility Reconstruction }
\label{vis_recon}
\begin{figure*}
    \centering
    \includegraphics[width=0.8\linewidth]{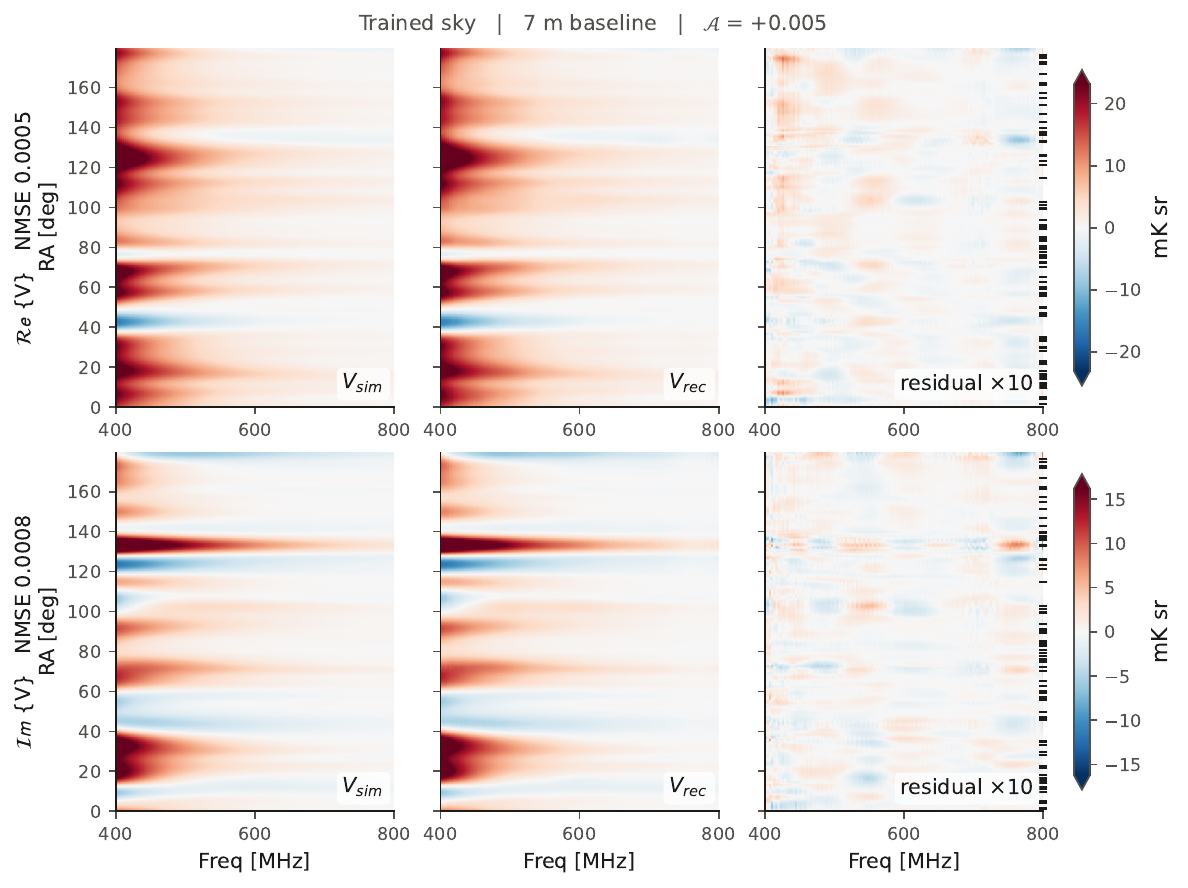}
    \caption{Real and Imaginary parts (in rows) of \textit{simulated}, \textit{reconstructed}, and their enhanced \textit{residual} difference visibilities (in columns) for the trained half-sky in $\mathrm{mK~sr}$ units for the shortest (7m) baseline for the instrument state $\mathscr{A} = +0.005$. The ticks on the right side of the residual panel indicate the RA slices held out for testing.}
    \label{fig:trained_recon}
\end{figure*}

\begin{figure*}
    \centering
    \includegraphics[width=0.8\linewidth]{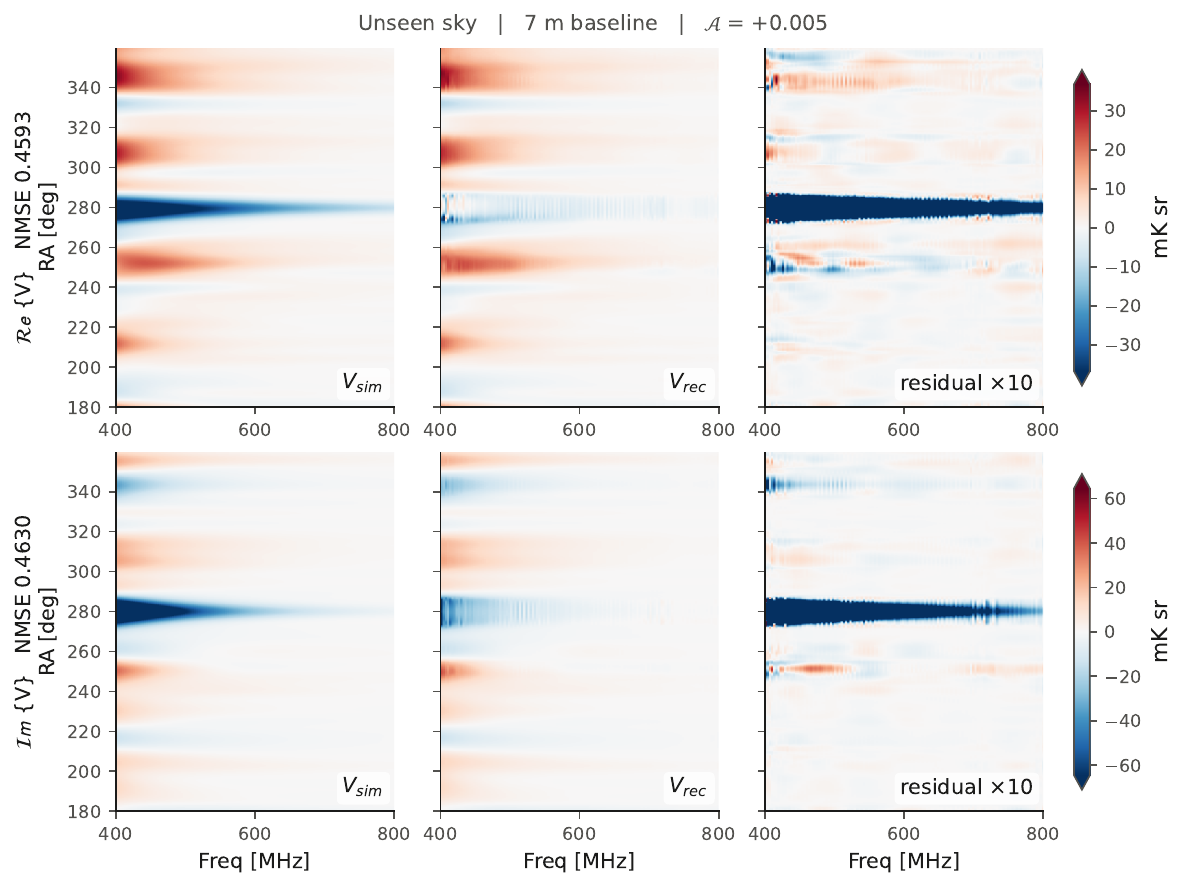}
    \caption{Same baseline and instrument state as in Fig.~\ref{fig:trained_recon}, but for the totally untrained half of the sky - the unused 512 RAs. }
    \label{fig:unseen_recon}
\end{figure*}

The visibility reconstruction error NMSE is governed by $L_{rec}$ during training, which was given equal weight like the two consistency losses. Although perfect reconstruction is not the core objective of this work, we acknowledge that reconstruction is highly significant for this framework to function in the long term. As we mentioned earlier, the entire work was done on one half of the sky, on which we trained and tested the model at low computational cost. In Fig. \ref{fig:trained_recon}, we show the real and imaginary parts of the \textit{simulated}, \textit{reconstructed}, and \textit{residual} visibilities for the trained sky half ($0^\circ \leq RA < 180^\circ$) for the shortest (7m) baseline and smallest ripple amplitude among the test seeds($\mathscr{A}=+0.005$). The instrument state is held out; the RA range
shown spans the sky half used in training, with the held-out pointings marked. The \textit{residual} (enhanced 10-fold) indicates that the reconstruction was satisfactory for this work, with NMSEs of 0.0005 for the real part and 0.0008 for the imaginary part, though still far from the level needed for cosmology (Sect.~\ref{discussion}). To stress test the model, we evaluate the same trained network on the unused, unseen half of the sky ($180^\circ \leq RA < 360^\circ$); we show the reconstruction and \textit{residual} in Fig. \ref{fig:unseen_recon} for the same baseline and $\mathscr{A}$. We can see that the reconstruction is good on part of the RAs but struggles particularly in the regions where V goes negative. The negative band near $\mathrm{RA} \approx 280^\circ$ is the Galactic plane transit, where a single bright off-zenith structure dominates the integral and carries its own fringe phase, in place of the diffuse near-zenith emission that keeps $V$ positive elsewhere. We also show how the reconstruction NMSE varies with $\mathscr{A}$ and baselines in Table \ref{tab:nmse_recon}, for three selected baselines (short, medium and long) and three instrument states, with the last one ($\mathscr{A}=-0.244$) being the extreme case. This extreme $\mathscr{A}$ has a z-score of $-2.44$, i.e., has only a $0.73\%$ chance of drawing a value this low; with 100 draws, about 0.7 such values are expected, so it lies in the extreme tail.

\begin{table}[ht]
\centering
\small
\begin{tabular}{c l S[table-format=1.4] S[table-format=1.4]}
\toprule
\textbf{$\mathscr{A}$} & \textbf{Baseline} & \textbf{Trained} & \textbf{Unseen} \\
\midrule
\multirow{3}{*}{$+0.005$} & \qty{7.0}{m}  & 0.0006 & 0.4620 \\
                          & \qty{28.0}{m} & 0.0033 & 0.0219 \\
                          & \qty{39.6}{m} & 0.0274 & 0.0274 \\
\midrule
\multirow{3}{*}{$+0.055$} & \qty{7.0}{m}  & 0.0015 & 0.4620 \\
                          & \qty{28.0}{m} & 0.0043 & 0.0239 \\
                          & \qty{39.6}{m} & 0.0272 & 0.0270 \\
\midrule
\multirow{3}{*}{$-0.244$} & \qty{7.0}{m}  & 0.0253 & 0.3585 \\
                          & \qty{28.0}{m} & 0.0270 & 0.0381 \\
                          & \qty{39.6}{m} & 0.0614 & 0.0640 \\
\bottomrule
\end{tabular}
\caption{Reconstruction errors for a short, medium, and long baseline for three different chromatic perturbation numbers $\mathscr{A}$, shown on the trained and unseen halves of the sky. Values are quoted to four significant figures; entries that appear equal differ beyond the quoted precision.}
\label{tab:nmse_recon}

\end{table}

\end{appendix}

\FloatBarrier 
\clearpage

\end{document}